\documentclass[conference]{IEEEtran}
\IEEEoverridecommandlockouts
\usepackage{cite}
\usepackage{amsmath,amssymb,amsfonts}
\usepackage{algorithmic}
\usepackage{graphicx}
\usepackage{textcomp}
\usepackage{xcolor}
\def\BibTeX{{\rm B\kern-.05em{\sc i\kern-.025em b}\kern-.08em
    T\kern-.1667em\lower.7ex\hbox{E}\kern-.125emX}}
\begin{document}

\title{A Resonant Tank Based Approach for Realizing ZPA in Inductive Power Transfer Systems \vspace{-0.5cm}
}

%\author{\IEEEauthorblockN{B Abhilash}
%\IEEEauthorblockA{\textit{Department of Electrical Engineering} \\
%\textit{IIT Madras}, Chennai, India. \\
%Email: }
%\and
%\IEEEauthorblockN{Arun Karuppaswamy B}
%\IEEEauthorblockA{\textit{Department of Electrical Engineering} \\
%\textit{IIT Madras}, Chennai, India.\\
%Email: akp@ee.iitm.ac.in}
%}

\author{B Abhilash \text{and} Arun Karuppaswamy B,~Department of Electrical Engineering, IIT Madras.\\ Email for correspondence: akp@ee.iitm.ac.in} 

\maketitle

\begin{abstract}
% Inductive power transfer (IPT) systems for electric vehicle (EV) applications need to achieve a load-independent constant current (CC) - constant voltage (CV) output for charging its battery. Further, zero phase angle (ZPA) is desirable to ensure a lower power rating requirement for the switching converter. Different resonant-tank based methods have been proposed for achieving CC-CV output without much mathematical effort, but achieving ZPA usually requires complicated equation manipulations which may not give useful physical insights. In this paper, a new perspective of seeing ZPA as equivalent to achieving load independent CC-CV output is proposed. Hence, the pre-existing methods like use of resonant tanks to achieve CC-CV output can be extended to achieve ZPA also. Also, the need of a separate method to achieve ZPA is eliminated. The proposed method is applied to a S-SP compensation topology and the obtained equations are compared with the results of another paper for verification. 

%-----------------------------------------------------------
A load-independent constant current (CC) - constant voltage (CV) output is an important requirement of inductive power transfer (IPT) systems for electric vehicle charging applications. Zero phase angle (ZPA) is also a desirable feature, to ensure a lower power rating requirement for the switching converter. CC and CV output along with ZPA can be achieved by using a suitable compensation topology. Equation manipulation techniques can be used for designing the compensation topology. But, it can be mathematically intensive especially for higher order topologies. To overcome this problem, resonant-tank based approaches are adopted in several works to achieve CC and CV conditions. However, equation-based approaches are depended upon either wholly or partly for realizing ZPA. This approach can be tedious and lacks physical insight. The proposed method extends resonant tank approach to achieve ZPA also, besides CC and CV. The need for a separate method to achieve ZPA is eliminated. Further, it simplifies the process in arriving at the constraints that ensure ZPA. As a sample validation, the proposed method is applied to a S-SP compensation topology. The CC-ZPA and CV-ZPA constraints for the S-SP topology are shown to be in line with the ones arrived at using an existing equation-based impedance approach. The simplicity of the proposed method can be observed from the sample validation. 
\end{abstract}

%The conditions for CC and/or CV along with ZPA, found using the proposed method, are in agreement with the constraints indicated in the literature for other topologies like SS, SP, PS, PP, double-sided LCL and double-sided LCC.


\section{Introduction}\label{intro}
%Electric vehicles (EVs) have several benefits such as lesser pollution and higher efficiency compared to internal combustion (IC) engine vehicles. As the usage of EVs grow, the need for public charging infrastructure becomes important. 

Static wireless inductive power transfer (IPT) for electric vehicle (EV) charging has several advantages compared to plug-in charging in terms of better safety, lesser maintenance requirement, etc. EVs generally use Li-ion batteries due to their high power densities. Constant current (CC)- constant voltage (CV) charging is the suitable charging method for Li-ion. Further, zero phase angle (ZPA) is desirable to ensure a lower power rating requirement for the switching converter. Some of the compensation topologies can be operated at two different frequencies to achieve CC and CV outputs along  with ZPA in both modes. Different methods have been proposed for identifying the constraints to achieve CC and CV along with ZPA \cite{RP3,RP28,RP31,RP48}. 

%The output port of the compensation topology which consists of a rectifier, a filter and the load can be modelled as an equivalent ac resistance $R_{ac}$ for these studies~\cite{RP31}.

\subsection{A brief survey of CC-CV and ZPA methods}
\begin{figure}[t]
\centerline{\includegraphics[scale=0.6]{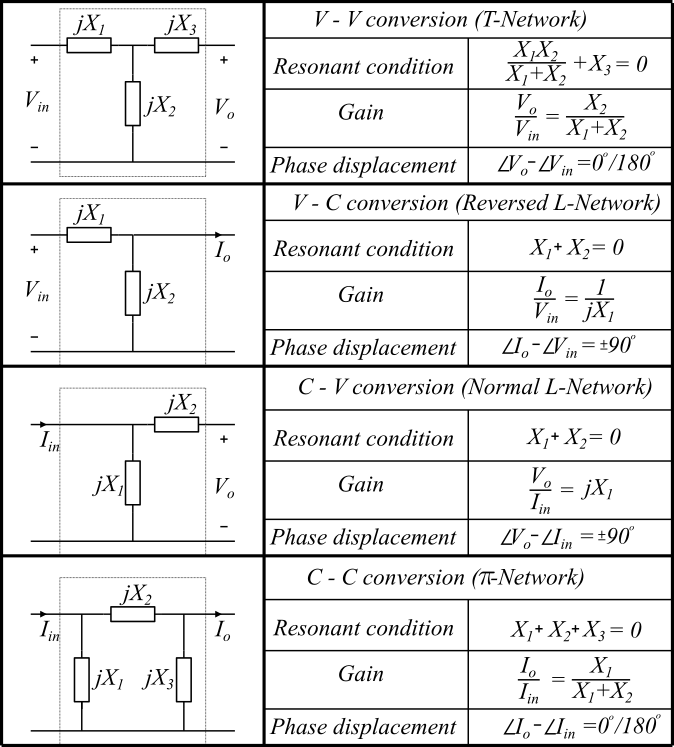}}
\caption{Basic resonant tanks and their properties~\cite{RP3}}
\label{restank}
\end{figure}

CC-CV output can be achieved using passive resonant networks. \cite{RP3} discusses that a T, reversed-L, normal-L and $\pi$ network can be used to achieve voltage-voltage (V-V), voltage-current (V-C), current-voltage (C-V) and current-current (C-C) conversions respectively by satisfying certain resonant conditions as given in Fig.~\ref{restank}. A compensation topology can be regarded as a cascade of these four basic resonant networks in order to understand its CC-CV functionality.

\begin{figure}[htbp]
\centerline{\includegraphics[scale=0.4]{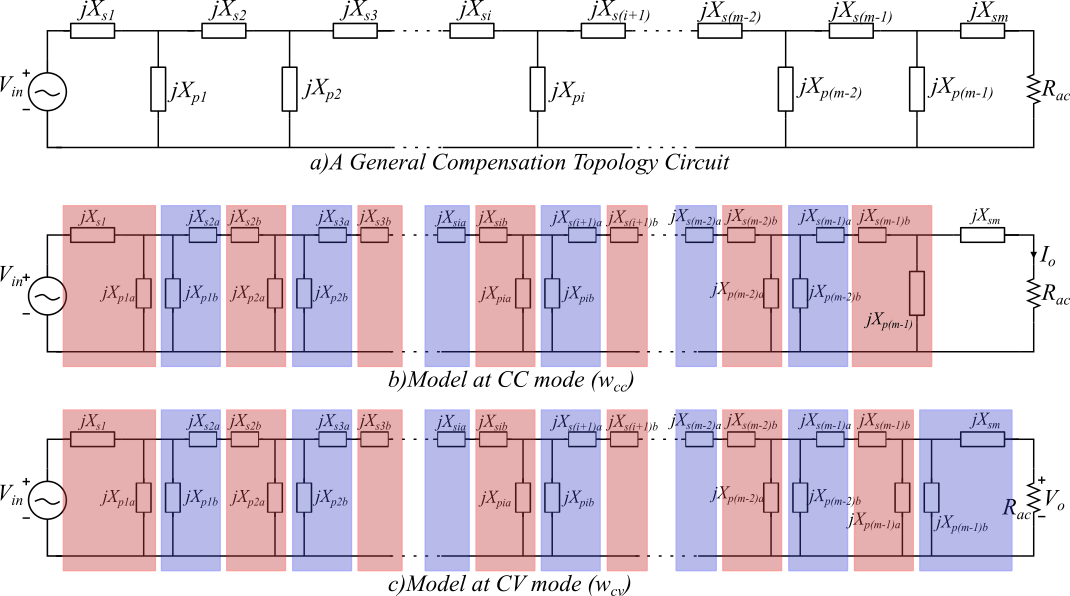}}
\caption{Unified model for an arbitrary compensation topology~\cite{RP28}}
\label{unifiedmodel}
\end{figure}

Further, \cite{RP28} points out that even a T- network and a pi-network can be regarded as a combination of  normal L-network and reversed L-network. Thus the entire compensation circuit can be disintegrated into normal and reversed L-networks cascaded alternatively as in Fig.~\ref{unifiedmodel} . A V-V conversion and a C-C conversion will require an even number of such L-networks, and a V-C conversion and a C-V conversion will require an odd number of L-networks.

Each L-network, normal or reversed, adds a phase of $\pm90^{\circ}$ between corresponding output and input quantities (Fig.~\ref{restank}). Since a V-V conversion requires an even number of resonant L-networks cascaded together, the output voltage ($V_o$) will always be shifted by a phase of $0^{\circ}$ or $180^{\circ}$ with respect to the input voltage ($V_{in}$). Similarly in a C-C conversion, the output and input currents are always displaced by $0^{\circ}$ or $180^{\circ}$. In C-V and V-C conversions, the corresponding output and input quantities are always phase displaced by $\pm90^{\circ}$ since an odd number of L-networks is used.

%Here I felt like continuity was not smooth When I read the paper for first time. Does adding this look better? "ref[RP28],etc have suggested different methods of achieving ZPA"  
For achieving ZPA, different methods are proposed by references \cite{RP3,RP28,RP31}. Reference \cite{RP3} discusses that ZPA can be achieved by making the imaginary part of the input impedance ($Z_{in}$) zero. This condition, $Im(Z_{in})=0$,  can be solved mathematically for any topology but it becomes tedious for higher order topologies. Reference \cite{RP28} proposes a general mathematical equation to achieve ZPA which works for any topology. The advantage of this method is that the impedances can be directly plugged in instead of deriving the condition from the first principles for every topology. However the equation is still complicated, especially for higher order topologies. Reference \cite{RP31} proposes the use of a T-network with their impedances satisfying a certain condition such that the input impedance becomes purely resistive. A compensation topology  can be seen as a cascade of such T-networks. This greatly reduces the computational effort required to solve a complicated mathematical equation. However, this method of achieving ZPA can be applied for only one mode, either CC or CV mode and the ZPA condition for the other mode has to be mathematically derived. Reference \cite{RP48} models the topology using gyrators to achieve CC-CV functionality and qualitatively comments on the sign of the input phase angle. 
%Ref[6] is not described here. Is that fine. ref[6] gives condition for CC/CV output by splitting the circuit into gyrators. It comments on input phase angle whether it is +ve/-Ve/zero, and thus comments on ZPA/ZVS.

In the proposed approach the ZPA condition is obtained by applying the resonant tank methods of \cite{RP3,RP28} which were till now used only to achieve CC-CV. Thus ZPA and CC-CV conditions can be seen from the same perspective and the need for a separate method to obtain ZPA is eliminated. Further, the proposed approach simplifies the process of arriving at the condition for ZPA. The equations that ensure ZPA can be written from the fundamental equations listed in Fig.~\ref{restank}. 
%The ZPA condition obtained by this method is in agreement with \cite{RP3,RP27,RP15} and for other topologies in literature as well.

\section{Principle of achieving ZPA}\label{principle}
% A CV mode with a voltage source input has to do V-V conversion using a compensation topology. The output voltage ($V_o$) and input current ($V_{in}$) will be displaced by $0^{\circ}/180^{\circ}$ as pointed out in section-\ref{intro}. 

% Now, if the compensation topology designed to do not only a V-V conversion but also a  C-C conversion, the output current($I_o$) and the input current($I_{in}$) will also be displaced by $0^{\circ}/180^{\circ}$. This ensures that the input phase angle ($\theta_{in}$) is the same as the output phase angle ($\theta_{out}$) as shown in \eqref{eq}.

% \begin{align}
%  \begin{array}{l}
% \angle V_{o} -\angle V_{in} =0^{\circ } /180^{\circ }\\
% \angle I_{o} -\angle I_{in} =0^{\circ } /180^{\circ }\\
% From\ ( 1) \ and\ ( 2) ,\\
% \angle V_{in} -\angle I_{in} =\angle V_{o} -\angle I_{o}\\
% ( or) \ \ \theta _{in} =\theta _{out}
% \end{array}\label{eq}
% \end{align}

An input voltage source requires a compensation topology with V-V conversion to achieve a CV mode. As pointed out in Section-\ref{intro} the phase of output voltage ($V_o$) and input voltage ($V_{in}$) will be displaced by $0^{\circ}/180^{\circ}$ (Fig.~\ref{restank}). \vspace{-0.5cm}

\begin{align}
\angle V_{o} -\angle V_{in} &=0^{\circ } /180^{\circ }  \label{eq1a} 
\end{align}

Now, if the compensation topology is designed to achieve C-C conversion along with V-V conversion, the phase of output current($I_o$) and the input current($I_{in}$) will also be displaced by $0^{\circ}/180^{\circ}$ (Fig.~\ref{restank}).\vspace{-0.6cm}

\begin{align}
\angle I_{o} -\angle I_{in} &=0^{\circ } /180^{\circ }  \label{eq1b} 
\end{align}

Then, we have either \vspace{-0.6cm}

\begin{align}
\angle V_{in} -\angle I_{in} &=\angle V_{o} -\angle I_{o} \nonumber\\
\theta _{in} &=\theta _{out} \label{eq1c}
\end{align}

or \vspace{-0.65cm}

\begin{align}
\angle V_{in} -\angle I_{in} &=(\angle V_{o} -\angle I_{o}) \pm 180^o \nonumber\\
\theta _{in} &=\theta _{out} \pm 180^o \label{eq1d}
\end{align}

The output port of the compensation topology which consists of a rectifier, a filter and the load can be modelled as an equivalent ac resistance $R_{ac}$~\cite{RP31}. The load being an equivalent resistance $R_{ac}$, $\theta_{out}$ will be $0^o$. If (\ref{eq1c}) is true, it would mean that power is being supplied to both the input and the output by the compensation network, which is not possible since the compensation network is just a combination of passive elements. This \textit{power flow constraint} requires \eqref{eq1c} to be true and not \eqref{eq1d}. So, $\theta_{in} = \theta_{out} = 0^o$ and ZPA is achieved.\\

Similarly in CC mode, if we ensure not only V-C conversion but also C-V conversion, $\theta_{in}$ is equal to $-\theta_{out}$ and ZPA can be achieved along with CC.

%as shown in \eqref{eq2} and ZPA can be achieved in CC mode.
\begin{align}
\angle I_{o} -\angle V_{in} &=\pm 90^{\circ }~~~\text{(V-C conversion)}\\
\angle V_{o} -\angle I_{in} &=\pm 90^{\circ }~~~\text{(C-V conversion)}\\
\angle V_{in} -\angle I_{in} &=-( \angle V_{o} -\angle I_{o}) \,, \ or \nonumber\\
\theta _{in} &=-\theta _{out} \label{eq2}
\end{align}

\section{Sample illustration of proposed approach}\label{application}
\begin{figure}[htbp]
\centerline{\includegraphics[scale=0.8] {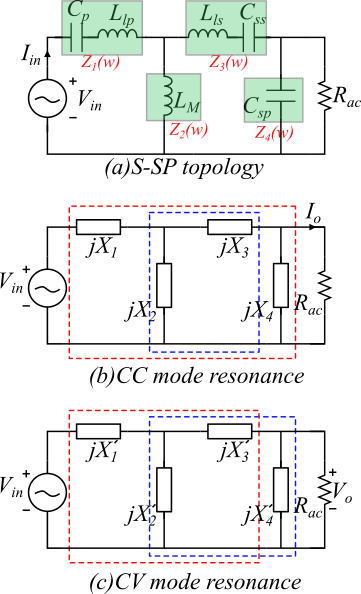}}
\caption{A S-SP topology and its resonant conditions in CC and CV modes. Red boxes indicate elements to be resonated for achieving CC or CV and blue boxes indicate the ones to be resonated for ZPA.}
\label{SSPmodelresonance}
\end{figure}

The proposed method for arriving at ZPA conditions is illustrated using a series – series parallel (S-SP) topology in this section. A S-SP topology has a series compensation capacitor ($C_p$) in the primary side, and a series capacitor ($C_{ss}$) and a parallel capacitor ($C_{sp}$) in the secondary side as shown in Fig~\ref{SSPmodelresonance}(a). $L_{lp}$, $L_{ls}$, and $L_{M}$ denote the primary and secondary leakage inductances and the magnetizing inductance of the wireless power transfer coil. The topology can be represented as in Fig~\ref{SSPmodelresonance}(a) with impedances $Z_1$ to $Z_4$ listed in \eqref{eq3a1} to \eqref{eq3a4}. Let $w_{cc}$ and $w_{cv}$ denote the CC mode and CV mode resonant frequencies. Let $jX_k$ represent the impedances $Z_k(jw_{cc})$ in CC mode and $jX_k^{'}$ represent the impedances $Z_k(jw_{cv})$ in CV mode as shown in \eqref{eq3b1} and \eqref{eq3b2}.

\begin{align}
Z_{1}(\omega) &=j\left( \omega L_{lp} -\frac{1}{\omega C_{p}}\right) \label{eq3a1} \\
Z_{2}(\omega) &=j( \omega L_M) \label{eq3a2} \\
Z_{3}(\omega) &=j\left( \omega L_{ls} -\frac{1}{\omega C_{ss}}\right)\label{eq3a3} \\
Z_{4}(\omega) &=j\left( -\frac{1}{\omega C_{sp}}\right) \label{eq3a4} \\
Z_{k}( \omega_{cc}) &=jX_{k}, k = 1,2,3,4  \label{eq3b1}\\
Z_{k}( \omega_{cv}) &=jX_{k}^{'}, k=1,2,3,4 \label{eq3b2}\end{align}

In Fig.~\ref{SSPmodelresonance}, the elements that need to be resonated to achieve CC/CV and the corresponding ZPA condition are shown in red and blue boxes respectively. 

\subsection{Conditions in CC mode}

\begin{figure}[htbp]
\centerline{\includegraphics[scale=0.8]{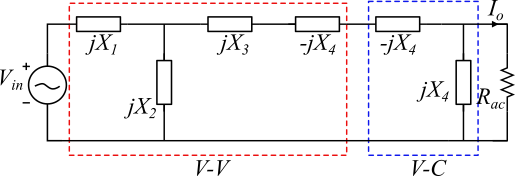}}
\caption{Derivation of constant current mode resonant condition for S-SP topology}
\label{CCmode}
\end{figure}

The CC mode can be obtained by resonating the components inside the red box in Fig.~\ref{SSPmodelresonance}(b). From Fig.~\ref{restank}, the required resonant condition is \eqref{eq4cca}. This can be understood by redrawing the circuit as shown in Fig.~\ref{CCmode}. The redrawn circuit is equivalent to Fig.~\ref{SSPmodelresonance}(b) and can be viewed as a cascaded combination of a V-V conversion and V-C conversion. The V-C conversion does not depend on the value of $X_4$ since any value of $X_4$ will satisfy the resonant condition for a reversed L-network given in Fig.~\ref{restank}.

%A slightly modified resonant tank approach is used for CC mode. Consider the circuit in Fig.~\ref{CCmode}(a). If the resonant condition \eqref{eq4cca} is ensured, this circuit will give a load independent output voltage $V_o = -V_{in}\cdot(X_3+X_4)/X_1 $. Now, if a notional voltage source equal to $V_o$ is connected to the output terminal it will not affect the node voltages and branch currents of the circuit in any way. So the circuit can be redrawn as in Fig~\ref{CCmode}(b) using substitution theorem. Now a source transformation can be done by converting the notional voltage source to notional current source as in Fig.~\ref{CCmode}(c). It is just a notional current source used for understanding the circuit behaviour, so even if it is removed and the output port is loaded with a resistance or short-circuited as shown in Fig.~\ref{CCmode}(d), the current will be same and independent of the load. Thus the S-SP topology satisfying the  resonant condition \eqref{eq4cca} does V-C conversion.

For ZPA condition, along with V-C (CC operation), C-V conversion is also required. Now, in order to achieve C-V conversion, the normal L-network consisting $X_2$ and $X_3$ can be resonated as \eqref{eq4ccb}. %The red and blue boxes in Fig.~\ref{SSPmodelresonance}(b) denote the CC and its ZPA resonant condition.
\begin{align}
CC:\ \frac{X_1 X_2}{X_{1} + X_2} + (X_{3} +X_{4}) &=0\label{eq4cca}\\
ZPA_{CC} :\ X_{2} +X_{3} &=0 \label{eq4ccb}
\end{align}

\subsection{Conditions in CV mode}
The CV mode can be obtained by resonating the T-network in Fig.~\ref{SSPmodelresonance} consisting $Z_1$, $Z_2$, $Z_3$ by ensuring the resonant condition \eqref{eq4cva}. For ZPA condition a C-C conversion is required. This can be done by resonating the $\pi$-network in Fig.~\ref{SSPmodelresonance} consisting $Z_2$, $Z_3$, $Z_4$ by ensuring \eqref{eq4cvb}. 

%The CV and corresponding ZPA condition are shown in red and blue boxes respectively, in Fig.~\ref{SSPmodelresonance}(c) 
% $1/Z_1+1/Z_2+1/Z_3=0$
\begin{align}
CV:\ \frac{X_1^{'} X_2^{'}}{X_{1}^{'} + X_2^{'}} + X_{3}^{'} &= 0 \label{eq4cva}\\
ZPA_{CV} :\ X_{2}^{'} +X_{3}^{'} +X_{4}^{'} &= 0 \label{eq4cvb}
\end{align}

\section{Equivalence to impedance approach}
\begin{figure}[htbp]
\centerline{\includegraphics[scale=0.8 ]{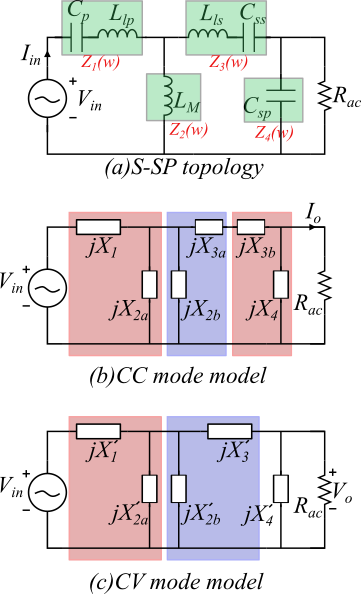}}
\caption{The unified model of S-SP topology used in \cite{RP28}}
\label{SSPmodelUnified}
\end{figure}

The equation based approaches in literature~\cite{RP15, RP27, RP28} typically equate the imaginary part of the input impedance of the compensation network to zero to arrive at the ZPA condition. This section compares the CC/CV and ZPA conditions derived in Section~\ref{application} for a S-SP topology to an existing impedance approach \cite{RP28} to show its equivalence. In~\cite{RP28}, called the unified model, the impedances are transformed as shown in Fig.~\ref{SSPmodelUnified} where the transformed impedances are given by \eqref{eq5a}, \eqref{eq5b} and \eqref{eq5c}. $w_{cc}$ and $w_{cv}$ denote the CC mode and CV mode resonant frequencies. %The equivalent variables in CC and CV modes are described by \eqref{eq5a} and \eqref{eq5b}, where $X_{2a}$, $X_{2b}$, $X_{3a}$, $X_{3b}$ are the transformed variables in CC mode and $X_{2a}^{'}$, $X_{2b}^{'}$ are the transformed variables in CV mode.

\begin{align}
\frac{X_{2a} X_{2b}}{X_{2a} +X_{2b}} &=X_{2} = \omega_{cc} L_M\label{eq5a} \\
X_{3a} +X_{3b} &= X_{3} = (\omega_{cc} L_{ls} - \frac{1}{\omega_{cc} C_{ss}})\label{eq5b} \\
\frac{X_{2a}^{'} X_{2b}^{'}}{X_{2a}^{'} +X_{2b}^{'}} &=X_{2}^{'} = \omega_{cv} L_M \label{eq5c}
\end{align}

The resonance condition derived in \cite{RP28} for CC output, ZPA at CC mode, CV output and ZPA at CV mode are given by : 
\begin{align}
CC:& \ X_{1} = -X_{2a} ,\ X_{2b} = -X_{3a} ,\ X_{3b} =-X_{4} \,, \label{eq6a} \\
ZPA_{CC} :& X_{2b} X_{3a} +( X_{2a} +X_{2b}) X_{3b} =0 \,, \label{eq6b} \\
CV:& \ X_{1}^{'} \ =-X_{2a}^{'} ,\ X_{2b}^{'} =-X_{3}^{'} \,, \label{eq6c} \\
ZPA_{CV} :& X_{2b}^{'} X_{3}^{'} +\left( X_{2a}^{'} +X_{2b}^{'}\right) X_{4}^{'} =0 \,. \label{eq6d}
\end{align}

\subsection{Verification in CC mode}
In CC mode, using \eqref{eq5b} and \eqref{eq6a}, the variables $X_{2a}$, $X_{2b}$, $X_{3a}$, $X_{3b}$ can be written as:
\begin{align}
    X_{2a} &=-X_{1} ,\ X_{2b} =-( X_{3} +X_{4}) , \nonumber\\
    \ X_{3a} &=X_{3} +X_{4} ,\ X_{3b} =-X_{4} \,. \label{eq7}
\end{align}

Substituting the above equation \eqref{eq7} in \eqref{eq5a} we get,

\begin{equation*}
    X_{2} =\frac{(-X_1)(-X_3-X_4)}{-X_1-X_3-X_4}
\end{equation*}

Rearranging, we get \eqref{eq8}, which is same as the CC resonant condition obtained in \eqref{eq4cca}.
\begin{equation}
\frac{X_1 X_2}{X_{1} + X_2} + (X_{3} +X_{4}) =0 \label{eq8}  
\end{equation}

Adding $X_{2a}(X_{3a}+X_{2b})$ to \eqref{eq6b} and rearranging, we get \eqref{eq9}. This does not alter the equation since $X_{3a}+X_{2b}$ is zero from \eqref{eq6a}.

\begin{equation}
    X_{2a} .X_{2b} \ +\ ( X_{3a} +X_{3b}) .( X_{2a} +X_{2b}) = 0 \label{eq9}
\end{equation}

Dividing by $(X_{2a}+X_{2b})$, we get \eqref{eq10} which is the same as the ZPA condition \eqref{eq4ccb}.

\begin{equation}
    X_{2} +X_{3} = 0 \label{eq10}
\end{equation}

% Dividing by $(X_{2a}+X_{2b})$ we get:
% \begin{equation}
%      \begin{array}{l}
% \frac{X_{2a} .X_{2b}}{( X_{2a} +X_{2b})} \ +\ ( X_{3a} +X_{3b}) =0\\
% ( or) \ X_{2} +X_{3} =0
% \end{array}
% \end{equation}
% which is same as the ZPA condition () obtained in this work.

\subsection{Verification in CV mode}
Similarly in CV mode, using \eqref{eq6c}, the transformed variables $X_{2a}^{'}$ and $X_{2b}^{'}$ can be written as
\begin{equation}
    X_{2a}^{'} =-X_{1}^{'} \,,\ and\ X_{2b}^{'} =-X_{3}^{'} \,. \label{eq11}
\end{equation}

Substituting the above equation in the expression for $X_{2}^{'}$ in \eqref{eq5c} and rearranging, we can get the same expression as \eqref{eq4cva} for the CV resonant condition.

%Substituting the above equation in the expression for $X_{2}^{'}$ in \eqref{eq5c} we get \eqref{eq12}, which is same as the CV resonant condition obtained in \eqref{eq4cva}.
%\begin{align}
%\frac{X_1^{'} X_3^{'}}{-(X_1^{'} + X_3^{'})} = X_2^{'} \, , \nonumber \\
%%\frac{1}{X_{2}^{'}} =\frac{1}{-X_{1}^{'}} +\frac{1}{-X_{3}^{'}} \,, \nonumber\\
%or, \frac{X_1^{'} X_2^{'}}{X_{1}^{'} + X_2^{'}} + X_{3}^{'} = 0 \label{eq12}
%%(or)\ \frac{1}{X_{1}^{'}} +\frac{1}{X_{2}^{'}} +\frac{1}{X_{3}^{'}} =0 \label{eq12}
%\end{align}

Adding $X_{2a}^{'}(X_3^{'}+X_{2b}^{'})$ to \eqref{eq6d} and rearranging, we get \eqref{eq13}. This does not alter the equation since ($X_3^{'}+X_{2b}^{'}$) is zero, from \eqref{eq6c}. Dividing \eqref{eq13} by ($X_{2a}^{'}+X_{2b}^{'}$), we get the ZPA condition which is the same as \eqref{eq4cvb}.
\begin{align}
    X_{2a}^{'} .X_{2b}^{'} \ +\ \left( X_{3}^{'} +X_{4}^{'}\right) .\left( X_{2a}^{'} +X_{2b}^{'}\right) =0 \label{eq13} 
%\frac{X_{2a}^{'} .X_{2b}^{'}}{\left( X_{2a}^{'} +X_{2b}^{'}\right)} \ +\ \left( X_{3}^{'} +X_{4}^{'}\right) =0\,, \nonumber\\
%( or) \ X_{2}^{'} +X_{3}^{'} +X_{4}^{'} =0 \label{eq14}
\end{align}

% Now, add $X_{2a}^{'}(X_3+X_{2b}^{'})$ to (14) and rearrange to get (). This will not cause any change because $X3+X_{2b}^{'}$ is zero from (17).
% \begin{equation}
%     X_{2a}^{'} .X_{2b}^{'} \ +\ \left( X_{3}^{'} +X_{4}^{'}\right) .\left( X_{2a}^{'} +X_{2b}^{'}\right) =0
% \end{equation}

% Dividing by $X_{2a}^{'}+X_{2b}^{'}$ we get:
% \begin{equation}
%      \begin{array}{l}
% \frac{X_{2a}^{'} .X_{2b}^{'}}{\left( X_{2a}^{'} +X_{2b}^{'}\right)} \ +\ \left( X_{3}^{'} +X_{4}^{'}\right) =0\\
% ( or) \ X_{2}^{'} +X_{3}^{'} +X_{4}^{'} =0
% \end{array}
% \end{equation}
% which is same as the ZPA condition ().

Thus, the conditions derived in this work is in agreement with the results obtained through equation based impedance approach outlined in \cite{RP28}. This is expected to be true for other impedance based methods available in literature as well since the angle between the input voltage and current being zero in the proposed approach is equivalent to an input impedance that is resistive.

%\section{Simulations}
%For the purpose of simulation, the design values by \cite{RP27} is considered and are given in Tab.~\ref{componentvalues} . Fig.~\ref{simulation} show the existence of CC and CV modes along with ZPA at 81.5kHz and 90kHz respectively under a changing load $R_{ac}$.

%\begin{table}[htbp]
%    \caption{Design values of \cite{RP27} for S-SP topology}
%    \begin{center}
%    \begin{tabular}{|c|c|c|c|}
%    \hline
%    \textbf{Parameters} & \textbf{Value (Unit)} & \textbf{Parameters} & \textbf{Value (Unit)} \\
%    \hline\hline
%    $L_p$ & $288\ \mu H$ & $C_p$ & $12.17\ nF$ \\
%    \hline
%    $L_s$ & $288\ \mu H$ & $C_ss$ & $13.25\ nF$ \\
%    \hline
%    $k$ & $0.14 $ & $C_sp$ & $60 nF$ \\
%    \hline
%    $f_{cc}$ & $81.5\ kHz$ & $f_{cv}$ & $90\ kHz$ \\
%    \hline
%    \end{tabular}
%    \label{componentvalues}
%    \end{center}
%\end{table}

%\begin{figure}[htbp]
%\centerline{\includegraphics[scale=0.25]{Conference/Images/SS_LIO_ZPA.pdf}}
%\caption{Simulation result depicting CC-CV and ZPA operation}
%\label{simulation}
%\end{figure}

\section{Conclusion}
In this paper a method to achieve ZPA condition using resonant tanks is presented. Thus CC-CV and ZPA conditions can be seen from the same lens and the need of two separate methods is eliminated. Further, the proposed approach largely simplifies the otherwise complicated equations required to arrive at the ZPA condition. The proposed method is applied for a S-SP compensation topology and the results are verified to be same as the results of \cite{RP28}. 

% Use the abbreviation ``Fig.~\ref{fig}''

% \begin{table}[htbp]
% \caption{Table Type Styles}
% \begin{center}
% \begin{tabular}{|c|c|c|c|}
% \hline
% \textbf{Table}&\multicolumn{3}{|c|}{\textbf{Table Column Head}} \\
% \cline{2-4} 
% \textbf{Head} & \textbf{\textit{Table column subhead}}& \textbf{\textit{Subhead}}& \textbf{\textit{Subhead}} \\
% \hline
% copy& More table copy$^{\mathrm{a}}$& &  \\
% \hline
% \multicolumn{4}{l}{$^{\mathrm{a}}$Sample of a Table footnote.}
% \end{tabular}
% \label{tab1}
% \end{center}
% \end{table}

% \begin{figure}[htbp]
% \centerline{\includegraphics{fig1.png}}
% \caption{Example of a figure caption.}
% \label{fig}
% \end{figure}

%\section*{Acknowledgment}

\end{document}